\documentclass[prl,superscriptaddress,showpacs,twocolumn,longbibliography]{revtex4-1}

\usepackage{hyperref}
\usepackage{color}
\usepackage[usenames,dvipsnames]{xcolor}
\usepackage{amsmath,amsthm,amssymb}
\usepackage{graphicx}
\usepackage{epsfig}
\usepackage{dcolumn}
\usepackage{bm}
\usepackage{mathrsfs}
\usepackage{multirow}
\usepackage[all]{xy}
\usepackage{pbox}
\usepackage{verbatim}
\usepackage{physics}

\def\(({\left(}
\def\)){\right)}
\def\[[{\left[}
\def\]]{\right]}

\newcommand{\be}{\begin{equation}}
\newcommand{\ee}{\end{equation}}
\newcommand{\ben}{\begin{eqnarray}}
\newcommand{\een}{\end{eqnarray}}
\newcommand{\beq}{\begin{equation}}
\newcommand{\eeq}{\end{equation}}

\newcommand{\la}{\langle}
\newcommand{\ra}{\rangle}

\begin{document}

\title{Entanglement statistics in Markovian open quantum systems: A matter of mutation and selection}

\author{Federico Carollo}
\affiliation{Institut f\"ur Theoretische Physik, Universit\"at T\"ubingen, Auf der Morgenstelle 14, 72076 T\"ubingen, Germany}
\author{Carlos P\'erez-Espigares}
\email{Corresponding Author: carlosperez@ugr.es}
\affiliation{Departamento de Electromagnetismo y F\'isica de la Materia, Universidad de Granada, Granada 18071, Spain}
\affiliation{Instituto Carlos I de F\'isica Te\'orica y Computacional, Universidad de Granada, Granada 18071, Spain}

\date{\today}

\begin{abstract}
Controlling dynamical fluctuations in open quantum systems is essential both for our  comprehension of quantum nonequilibrium behaviour and for its possible application  in near-term quantum technologies.  However, understanding these  fluctuations is extremely challenging due, to a large extent, to a lack of efficient important sampling methods for quantum systems. Here, we devise a unified framework --based on population-dynamics methods-- for the evaluation of the full probability distribution of generic time-integrated observables in Markovian quantum jump processes. These include quantities carrying information about genuine quantum features, such as quantum superposition or entanglement, not accessible with existing numerical techniques. The algorithm we propose provides dynamical free-energy and entropy functionals which, akin to their equilibrium counterpart, permit to unveil intriguing phase-transition behaviour in quantum trajectories. We discuss some applications and further disclose coexistence and hysteresis, between a highly entangled phase and a low entangled one, in large fluctuations of a strongly interacting few-body system.
\end{abstract}

\maketitle

Fluctuations, even those well far from the average, play a fundamental role both in classical and quantum systems. In equilibrium, for instance, it is necessary to know the probability of {\emph {every}} microscopic configuration in order to predict, via the singularities in the partition function, the existence of phase transitions. This exceptional result of standard statistical mechanics, however, breaks down as soon as we drive the system far from equilibrium. In this scenario, strong spatio-temporal correlations that are absent in equilibrium emerge \cite{Linkenkaer-Hansen1370,Levina:2007aa,Choi:2017aa,Zhang:2017aa,Xiong:2019aa} and stationary properties alone are not sufficient to understand the physics of the system, being required to study its trajectories and their statistical properties. To this aim, a thermodynamic formalism for trajectories  --built on large deviations \cite{touchette09a}-- has been devised \cite{ruelle_2004,PhysRevLett.74.2694,PhysRevE.52.3525,lecomte07c,garrahan09a}, leading to important breakthroughs in the nonequilibrium realm \cite{merolle05a,garrahan07a,hedges09a,bertini15a,lesanovsky13b}. Nonetheless, its application to the exploration of dynamical fluctuations of genuine quantum features, such as entanglement, has been elusive so far.

Extensively investigated in equilibrium and in unitary dynamics \cite{Osterloh:2002aa,RevModPhys.82.277,Jurcevic:2014aa,Alba201703516,PhysRevLett.121.170602}, entanglement is a crucial resource for quantum information and metrology \cite{Giovannetti1330,Giovannetti:2011aa,RevModPhys.81.865}, and further represents a  measure of complexity for the many-body wave-function. Its notion has led to important advances in subjects ranging from thermalization or quantum thermodynamics \cite{Kaufman794,Vedral:2014aa} to black holes and cosmology \cite{Steinhauer:2016aa,Mart_n_Mart_nez_2014}. Recently, the dynamics of entanglement in open systems and in stochastic pure state evolutions has been the focus of intense research \cite{PhysRevX.9.021007}. From the analysis of  trajectories, it has been possible to observe nonequilibrium entanglement phase transitions \cite{PhysRevB.99.224307,PhysRevB.98.205136,PhysRevX.9.031009} and to develop a dissipative quasi-particle picture in non-interacting systems \cite{10.21468/SciPostPhys.7.2.024}. However, fundamental questions like ``What is the probability of observing a given entanglement fluctuation?" or ``Is it possible to observe phase transitions, analogous to those in equilibrium, in entanglement fluctuations?" have, as yet, not been fully addressed. 

\begin{figure}[t]
\centering
\includegraphics[scale=0.28]{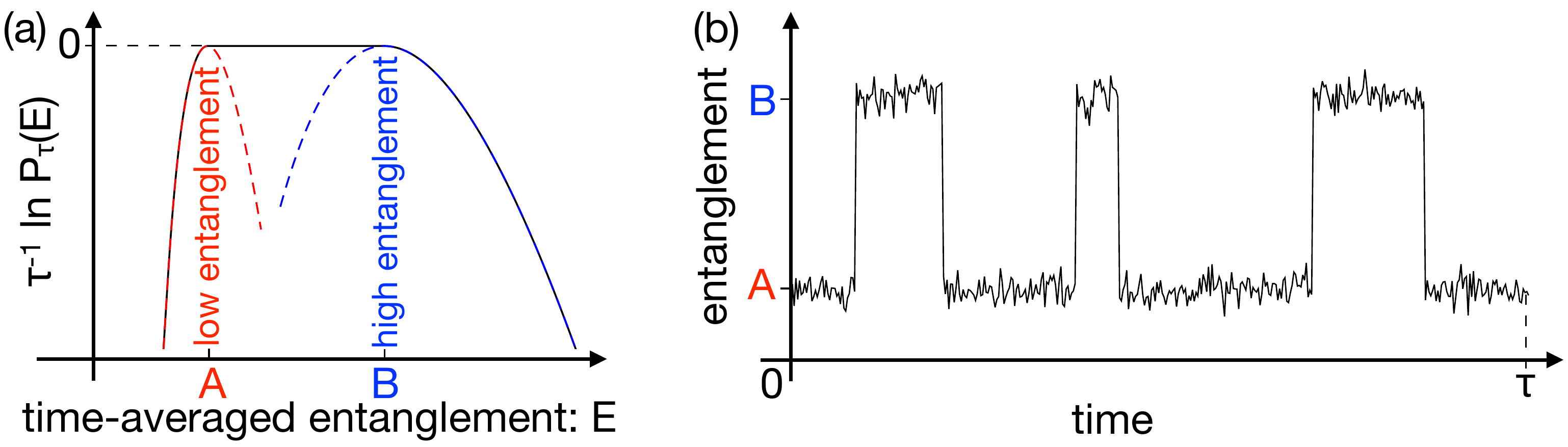}
\caption{{\bf Sketch of Coexistence of  Dynamical Regimes.} (a) Statistics of  time-averaged entanglement showing bimodality of two regimes, a high and a low entanglement one. Fluctuations with time-averaged entanglement between the typical values of the two regimes are characterised by alternating behaviour in time, as shown by a representative trajectory in (b). If the phases are equally likely, then all possible intermediate values for the time-averaged entanglement also are. }
\label{Fig1}
\end{figure} 

Here, borrowing ideas from classical nonequilibrium \cite{giardina06a,giardina11a}, we develop a unified framework for the evaluation of the  probability distribution of dynamical fluctuations of arbitrary time-integrated observables in quantum jump processes \cite{plenio98a,gardiner00a}. This approach thus permits to uncover the statistics of a number of observables not accessible with state-of-the-art large deviation algorithms \cite{garrahan10a,carollo18b,carollo19a}. To illustrate its potential, we derive the  dynamical counterpart of equilibrium free-energy and entropy functionals for entanglement \cite{PhysRevLett.101.050502,PhysRevLett.104.110501} and quantum coherence. We further unveil critical phenomena with coexistence --as sketched in Fig.~\ref{Fig1}-- and hysteresis between a high and a low entanglement phase, in fluctuations of strongly interacting few-body systems. 
\\

\noindent {\bf \em Quantum trajectories-- } We  focus on open quantum systems described by quantum jump processes \cite{plenio98a,gardiner00a}. The average  quantum state $\rho_t$ evolves through the Lindblad master equation \cite{lindblad76a,gorini1976}
\begin{equation}
\frac{d\rho_t}{dt}=-iH_{\rm eff}\rho_t+i\rho_t H_{\rm eff}^\dagger+\sum_{\mu=1}^{N_{\rm J}} 
J_\mu \rho_t J_\mu^\dagger\, ,
\label{Lindblad}
\end{equation}
where $H_{\rm eff}=H-i/2\sum_\mu J^\dagger_\mu J_\mu$ is the effective Hamiltonian, with $H$ being the Hamiltonian of the system.  $N_{\rm J}$ stands for the number of jump operators $J_{\mu}$ encoding the system-bath interaction. However, the process is stochastic and much more information than in the average dynamics \eqref{Lindblad} is contained in quantum trajectories. These are dynamical realisations, in which an initial pure state follows an overall deterministic evolution with $H_{\rm eff}$, punctuated by stochastic jumps  at random times. We exploit a discrete-time approximation, where a trajectory of length  $\tau$ is divided into $N$ infinitesimal time-steps $\delta t=\tau/N\to0$. At each time $t=k\delta t$, the pure state $|\psi_t\rangle$  either jumps to $|\psi_{t+\delta t}\rangle=J_\mu |\psi_t\rangle/\left\|J_\mu|\psi_t\rangle\right\|$, for $\mu=1,2,\dots N_{\rm J}$, with probability $p_\mu=\delta t\, \langle \psi_t|J^\dagger_\mu J_\mu|\psi_t\rangle$, or evolves under $H_{\rm eff}$ to
$|\psi_{t+\delta t}\rangle=e^{-i\delta t H_{\rm eff}}|\psi_t\rangle/ \left\|e^{-i\delta t H_{\rm eff}}|\psi_t\rangle\right\|$,
with probability $p_0=1-\sum_{\mu=1}^{N_{\rm J}}p_\mu$. 
\\

\noindent {\bf \em Statistics of time-integrated observables-- }
We now briefly present the formalism that allows us to derive the statistics of time-integrated observables in quantum jump processes.
To this end, we notice that a quantum trajectory, denoted by $\omega_\tau$, consists of a sequence of pure states $\omega_\tau=\{ |\psi_0\rangle, |\psi_1\rangle, \dots |\psi_{N}\rangle \}$, where $|\psi_k \rangle$ is short-hand for $|\psi_{t=k\, \delta t}\rangle$, whose
probability is given by 
\be
P[\omega_\tau]=p_{|\psi_0\rangle}p_{|\psi_0\rangle \to |\psi_1\rangle}\dots p_{|\psi_{N-1}\rangle \to |\psi_{N}\rangle}\, .
\label{probtraj}
\ee
Here $p_{|\psi_0\ra}$ is the probability of the initial (pure) state while $p_{|\psi_k\rangle \to |\psi_{k+1}\rangle}$ is the transition probability at $t=k\delta t$, corresponding either to $p_{\mu}$ or to $p_0$ depending on whether the state jumps with $J_\mu$ or evolves under $H_{\rm eff}$. The full set of trajectories and associated probabilities completely characterises the process. However, this information is often intractable and experimentally irrelevant, the focus thus being on  the statistics of one or several time-integrated observables $O_\tau$. Each of these  can be expressed as a functional over trajectories, $O_\tau=\mathcal{O}[\omega_\tau]$ and, in the discrete-time approximation, written as 
\be
\mathcal{O}[\omega_\tau]=\sum_{k=0}^{N-1}\alpha\left(|\psi_k\rangle,|\psi_{k+1}\rangle\right)\, ,
\label{obstraj}
\ee
where $\alpha$ is an observable-dependent function describing the discrete increment of $O_\tau$. As such, the probability distribution of $O$ can be obtained by contraction from that of trajectories $P_{\tau}(O)=\sum_{\omega_\tau} P[\omega_\tau]\, \delta(O-\mathcal{O}[\omega_\tau])$. For long times $\tau$, the above probability obeys a large deviation principle $P_{\tau}(O)\sim e^{\tau\, \varphi(o)}$, with $o=\tau^{-1}O$ being the time-averaged value of the observable and $\varphi\le0$ the so-called rate function \cite{touchette09a}. This has the flavour of an entropy functional for nonequilibrium systems, with the duration of the trajectory $\tau$ playing the role of the volume and the typical outcome for the observable, $\la o \ra$, maximizing the value of the entropy, $\varphi(\la o \ra)=0$. 

\begin{figure}[t!]
\hspace{-0.2cm}\includegraphics[scale=0.24]{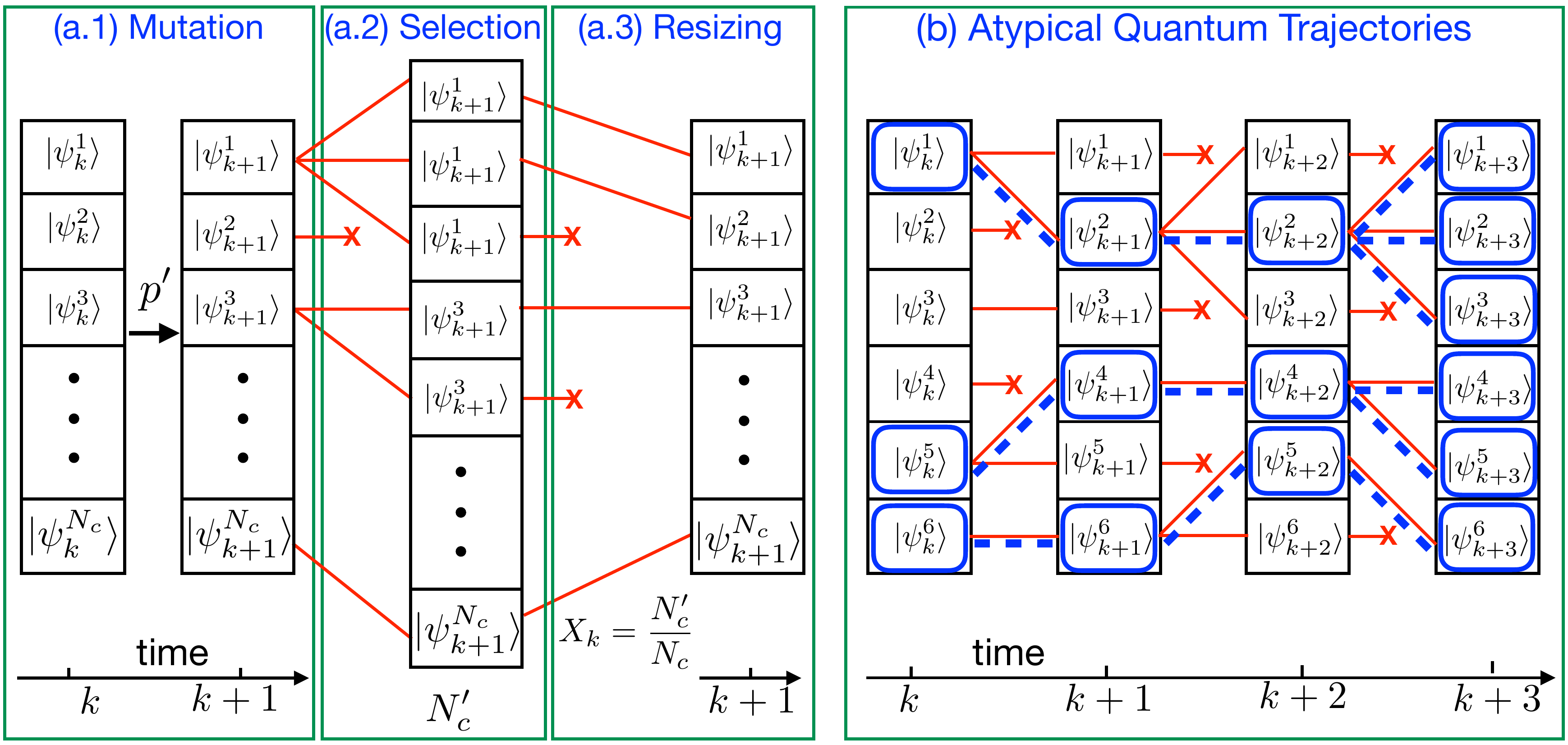}\\
\caption{{\bf Quantum Cloning Algorithm.}
(a) Sketch of the updating scheme for a time-step. (a.1) Clones evolve  independently  according to their biased probabilities $p'$. (a.2) Each clone $i$ is reproduced or pruned with average $Y_{|\psi_k^i\rangle}$. (a.3) This changes the size of the  population which  is resized to its original value by randomly pruning (or replicating) clones. (b) Illustration of six atypical trajectories that are retrieved by tracing backward clones which survived up to the last time-step.
}\label{Fig2}
\end{figure}

Dual to this microcanonical picture, one may introduce a canonical ensemble of trajectories
$P_s[\omega_\tau]=e^{-s\, \mathcal{O}[\omega_\tau]}P[\omega_\tau]/\mathcal{Z}_\tau(s)$ --with $\mathcal{Z}_\tau(s)$ being the (normalising) partition function--
through a temperature-like parameter $s$. In this  ensemble the microcanonical constraint is softened and  only time-averages are controlled by $s$. For large $\tau$, canonical and micro-canonical  pictures are equivalent \cite{chetrite13a}, and the dynamical partition function 
$
\mathcal{Z}_\tau(s)=\sum_{\omega_\tau} e^{-s\, \mathcal{O}[\omega_\tau]}P[\omega_\tau]\, 
$,
behaves as $\mathcal{Z}_\tau(s)\sim e^{\tau\, \theta(s)}$ where $\theta(s)$ is the scaled cumulant generating function of the observable $o$. The function $\theta(s)$, which is the Legendre transform of $\varphi(o)$, plays a role akin to (minus) the  free-energy and completely characterises the statistics of $o$.  Analogously to equilibrium settings,  singular points in $\theta$ and $\varphi$ may disclose critical behaviour in the space of quantum trajectories \cite{touchette09a,garrahan10a}. 
\\

 \begin{figure*}[t]
\centering
\includegraphics[scale=0.55]{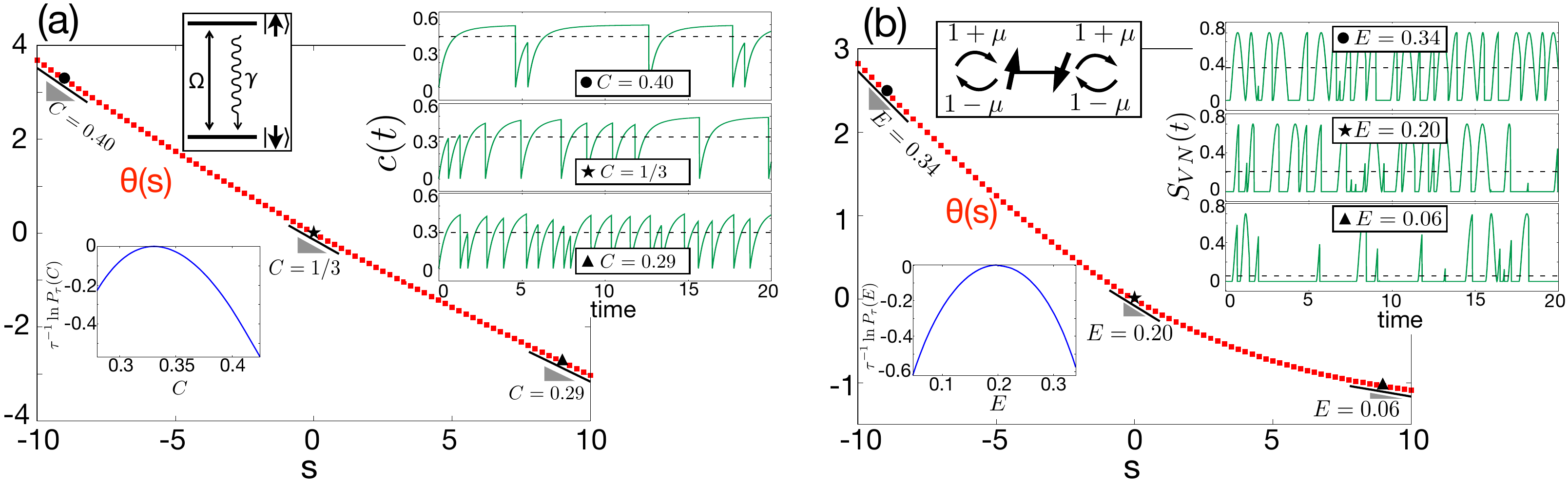}
\caption{{\bf Statistics of coherence and entanglement.} (a) Dynamical free-energy $\theta(s)$ and probability $P_\tau(C)$ for the time-averaged coherence in the laser-driven two-level atom with emission decay sketched in the panel (for $\Omega=1$ and $\gamma=4$). Coherence measures the quantum superposition between the classical configurations $\ket{\uparrow}$ and $\ket{\downarrow}$. The function $\theta(s)$ is smooth indicating the existence of a single dynamical phase. Jumps reset coherence to zero while the deterministic evolution generates superposition. The average coherence is $C=1/3$. Large fluctuations characterised by larger (smaller) value of $C$ are simply generated by larger (smaller)  times between jumps. (b) Dynamical free-energy $\theta(s)$ and probability $P_\tau(E)$ for the time-averaged entanglement entropy $E$ in a boundary-driven two-site non-interacting chain with $\mu=0.5$. There exists as well a single dynamical phase. The deterministic dynamics can generate entanglement; however, any boundary event then collapses the system into a separable state, which does not evolve until a further jump occurs. Fluctuations with larger (smaller) than typical entanglement are characterised by longer (shorter) survival times of the entangled state and shorter (longer) survival times of separable states.}
\label{Fig3}
\end{figure*} 

\noindent {\bf \em Numerical method-- } Our approach consists in  computing $\theta(s)$ by exploiting a quantum version of the so-called cloning algorithm \cite{giardina06a,giardina11a}. If we insert Eqs.\eqref{probtraj},\eqref{obstraj} into $\mathcal{Z}_\tau(s)$, we get
\be
\mathcal{Z}_\tau(s)=\sum_{\omega_\tau} p_{|\psi_0\rangle} Y_{|\psi_0\rangle}p'_{|\psi_0\rangle \to |\psi_1\rangle}\dots Y_{|\psi_{N-1}\rangle}p'_{|\psi_{N-1}\rangle \to |\psi_{N}\rangle}\, ,
\label{dynPF_v1}
\ee
with $p'_{|\psi_{k}\rangle \to |\psi_{k+1}\rangle}=e^{-s\alpha(|\psi_{k}\rangle, |\psi_{k+1}\rangle)} p_{|\psi_{k}\rangle \to |\psi_{k+1}\rangle}/Y_{|\psi_k\rangle}$ and $Y_{|\psi_k\rangle}$ being the  normalisation factor.

The partition function written as in Eq.~\eqref{dynPF_v1} has now a clear computational interpretation, as illustrated in Fig.~\ref{Fig2}(a). We consider a number of copies of the system, $N_c$, which independently evolve or ``{\emph{mutate}}" according to the biased stochastic dynamics $p'_{|\psi^i_{k}\rangle \to |\psi^i_{k+1}\rangle}$, for $i=1,\dots,N_c$, see Fig.~\ref{Fig2}(a.1). After this mutation step, a ``\emph{selection}"  takes place and every clone is either stochastically reproduced by an integer number of copies or pruned with average $Y_{|\psi^i_k\ra}$, as displayed in Fig.~\ref{Fig2}(a.2). 
This step changes the evolving population to the value $N'_c$. In order to avoid diverging or vanishing populations during the dynamics, this  must be resized, as shown in Fig~\ref{Fig2}(a.3), by killing or replicating randomly selected clones. This terminates the evolution of clones over a time-step. 
With this procedure, the dynamical free-energy is $\theta(s)=\tau^{-1}\sum_{k=1}^N\ln(X_k)$, with $X_k=N'_c/N_c$ being the resizing factor at step $k$ (see S1 in Supplemental Material \cite{SM} for a detailed discussion of the algorithm). 

In addition to the full statistics obtained via  $\theta(s)$, clones surviving up to the last time-step provide a representative sample of the quantum trajectories 
realising the fluctuation $ o=-\theta'(s)$, {\it c.f.} Fig.~\ref{Fig2}(b). In the presence of critical phenomena in trajectory-space \cite{perez-espigares19a}, 
the analysis of these samples can shed light on the properties of the dynamical phases of the system.
\\

\begin{figure*}[t]
\centering
\includegraphics[scale=0.52]{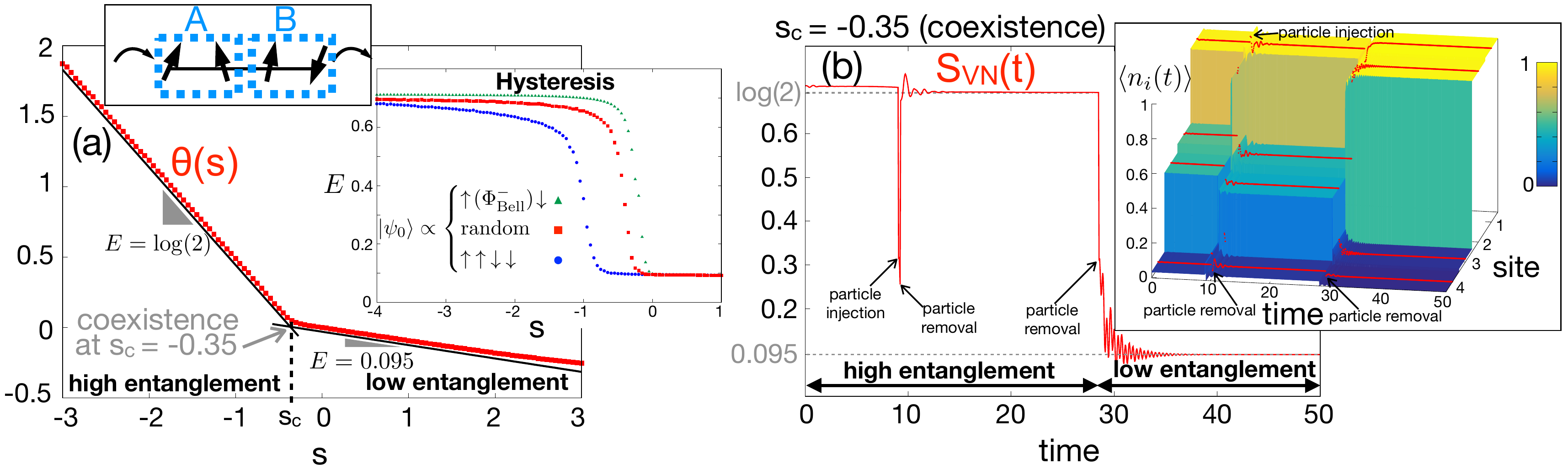}
\caption{ {\bf Dynamical coexistence and hysteresis in entanglement fluctuations.} (a) Dynamical free-energy $\theta(s)$ with random initial conditions for strongly interacting boundary-driven XXZ chain with $L=4$ spins and $\mu=1$, as sketched in the panel. The sudden change in slope signals the existence of two dynamical regimes, a high ($E\approx \log 2$) and a low entanglement ($E\approx 0.095$) one. In the inset we show hysteresis behaviour of entanglement fluctuations. For separable initial conditions the entanglement crossover takes place at a  larger value of $|s|$ than for entangled initial conditions. (b) Trajectory of the half-chain entanglement Von Neumann entropy $S_{VN}(t)$ at the dynamical crossover point $s_c=-0.35$, for random initial conditions. Analogously to what happens in equilibrium, we observe coexistence in time of values of entanglement characteristic of the two phases. The inset shows the corresponding trajectory of the average occupation number: in the entangled phase we observe superposition of states in the central spins, while the low entanglement phase is close to a classical configuration.}
\label{Fig4}
\end{figure*} 

\noindent {\bf \em Coherence and entanglement statistics-- } Within this framework it is possible to explore completely novel aspects about dynamical fluctuations of genuine quantum features. Indeed, the algorithm is not restricted to jump-rate observables like currents or activities (see S2 in \cite{SM} for examples of these observables) but it is rather {\it universal}, in that it can be used for generic time-integrated observables once the proper increment $\alpha$ is identified. Thus to unveil the statistics of a function of the state $f(|\psi\rangle)$ --where $f$ must be understood as a function of some or all elements of the vector $|\psi\rangle$-- the increment can be written as $\alpha(|\psi_k\rangle,|\psi_{k+1}\rangle)=(f(|\psi_k\rangle)+f(|\psi_{k+1}\rangle))\delta t/2$.
The sum in equation \eqref{obstraj} provides, for $\delta t\to 0$, a good approximation to the  observable $O_\tau=\int_0^\tau f(|\psi_t\rangle) dt$. Notice that tilted Lindblad  techniques \cite{garrahan10a,carollo18b} cannot provide the statistics of such observables \cite{carollo19a}.

In Fig.~\ref{Fig3} we report on the characterisation of the statistics of time-averaged coherence and entanglement in paradigmatic few-body systems. For coherences, defined as $C=\tau^{-1}\int_0^{\tau} dt\, c(t)$, we consider a two-level atom with $J_1=\sqrt{\gamma} \sigma_-$ and $H=\Omega (\sigma_{-}+\sigma_+)$, with $\sigma_\pm$ being the ladder operators \cite{garrahan10a}. Here, $c(t)$ is the modulus of the off-diagonal terms of the stochastic state $|\psi_t\rangle \langle \psi_t|$ and represents a measure of quantum superposition (see S3 in \cite{SM} for more details). The dynamical free-energy, see Fig.~\ref{Fig3}(a), is smooth in the temperature-like parameter $s$ indicating that trajectories are dominated by a single dynamical phase. The deterministic dynamics in between jumps builds up coherence until a jump --or photon emission-- resets the system into the de-excited (classical) state. Thus, fluctuations with large time-integrated coherence are characterised by smaller than typical jump rates, as displayed in Fig.~\ref{Fig3}(a).

We then consider the statistics of the time-averaged half-system entanglement --Von Neumann-- entropy, $E=\tau^{-1}\int_0^{\tau}S_{VN}(t)dt$, in a non-interacting two-site boundary-driven chain \cite{znidaric14a} [see Fig.~\ref{Fig3}(b)] (see S4 in \cite{SM} for more details). The model is described by equation \eqref{Lindblad}, with $J_{1/2}=\sqrt{\gamma_{\pm}}\sigma_{\pm}^{(1)}$ injecting/ejecting a particle in the first site and $J_{3/4}=\sqrt{\gamma_{\mp}}\sigma_{\pm}^{(L)}$ in the last. Rates $\gamma_{\pm}=1\pm \mu$ depend on the driving parameter $\mu$, and $
H=\sum_{k=1}^{L-1}\left(\sigma_-^{(k)}\sigma_+^{(k+1)}+\sigma_+^{(k)}\sigma_-^{(k+1)}\right)+\Delta\sum_{k=1}^{L-1}\sigma_z^{(k)}\sigma_z^{(k+1)}$, with $\Delta$ accounting for interactions. 

In Fig.~\ref{Fig3}(b), we observe a smooth entanglement dynamical free-energy. The dynamics is made of boundary jump events which destroy entanglement, if present, by collapsing the state onto classical configurations. These are time-invariant under $H_{\rm eff}$ so they survive until a further boundary event occurs. High (low) entanglement fluctuations are generated by dynamical realisations with small (large) fractions of time spent in classical configurations, as shown by representative quantum trajectories.

It is worth remarking that generic quantum jump processes --contrary to classical ones-- do not visit the whole space of admissible pure states \cite{carollo19a}. Initial conditions thus play a crucial role. For instance, a process could generate a large entanglement fluctuation from a highly entangled initial state simply avoiding jump events, in order not to leave a favorable dynamical regime which, once left, could not be reproduced. This however does not reflect the {\it entangling capability} of the process: to unambiguously characterise the entanglement statistics of the process itself, one must start from separable states, as in Fig.~\ref{Fig3}, or investigate different initial conditions, as we do in what follows. 
\\

\noindent {\bf \em Dynamical crossover, coexistence and hysteresis between high and low entanglement phases--} The dynamical free-energy of entanglement may also feature abrupt changes associated with different fluctuations. Analogously to what happens in equilibrium, this is the signal of transition-like critical behaviour in trajectories. 

In Fig.~\ref{Fig4}(a) we report on a sudden change of the dynamical free energy $\theta(s)$ of the half-chain entanglement entropy, in a strongly interacting ($\Delta=4$) few-body boundary-driven chain \cite{PhysRevLett.107.137201} (see S4 in \cite{SM} for more details). This witnesses the existence of two distinct dynamical regimes,  reminiscent of first-order phase transitions as sketched in Fig.~\ref{Fig1}. In particular, these two regimes correspond to a high and a low entanglement phase. Critical fluctuations of entanglement with values in between these two regimes are characterised by coexistence of both phases, as displayed by a representative trajectory in Fig.~\ref{Fig4}(b). Here we see how a highly entangled phase  ($S_{VN}(t)\approx\log2$), with central spins having $~76\%$ projection on the Bell state $|\Phi_{\rm Bell}^-\rangle \propto \ket{\uparrow \uparrow} -\ket{\downarrow \downarrow}$, deterministically evolves until quantum jumps take place and drive the system into a lower entanglement phase ($S_{VN}(t)\approx 0.095$), with a $~98\%$ projection on the classical configuration $\ket{\uparrow \uparrow\downarrow\downarrow}$. This intermittent behavior manifests as well in the average occupation number --see inset to Fig.~\ref{Fig4}(b). 

The function $\theta(s)$ has been obtained starting from a random initial condition for each clone. Once we have approximately established the reference states for the two dynamical regimes, we can verify the robustness of this abrupt crossover against initial conditions. 
In the inset to Fig.~\ref{Fig4}(a), we show the time-averaged entanglement for different initial conditions. Remarkably, hysteresis behaviour emerges in entanglement fluctuations. When starting from a separable state the crossover takes place at a larger finite value of $|s|$ --meaning that the large entanglement phase is less likely observed-- than when the initial state is $\ket{\uparrow \Phi^-_{\rm Bell}\downarrow}$, closer to the reference state for the highly entangled phase. Since the sudden crossover persists when starting from a fully separable state, the entanglement critical dynamics is intrinsic of the quantum process and should not be associated with the entanglement of initial conditions. 
\\

\noindent {\bf \em Outlook-- } We have introduced a general numerical framework where statistical properties of arbitrary time-integrated observables of quantum jump trajectories can be explored. This approach paves the way to the study of the fluctuating behaviour of truly quantum features. 

Further interesting applications of this algorithm are the study of genuine many-body dynamical phases and the characterisation of the wave-function complexity in quantum stochastic processes. To this end, one needs to combine it with approximate representations of many-body states in quantum trajectories. As we show in \cite{SM}, we can straightforwardly combine the quantum cloning with matrix product state (MPS) representations of quantum states \cite{Perez-Garcia:2007,PAECKEL2019167998}. Indeed, one just needs to run several trajectories using MPS methods (as is done, e.g.,  in \cite{Gillman_2019}) and to apply to such evolving population of clones the selection dynamics explained in our paper. The preliminary results in \cite{SM}, obtained through a very simple MPS algorithm, show indeed promising applications of our quantum cloning algorithm to many-body systems also in situations with an extensive number of jump operators. 

\begin{acknowledgments}
The research leading to these results has received funding from the European Union's Horizon 2020 research and innovation programme under the Marie Sklodowska-Curie Cofund Programme Athenea3I Grant Agreement No. 754446, from the European Regional Development Fund, Junta de Andaluc\'ia-Consejer\'ia de Econom\'ia y Conocimiento, Ref. A-FQM-175-UGR18 and from EPSRC Grant No. EP/N03404X/1. F.C. acknowledges support through a Teach@T\"ubingen Fellowship.  We are also grateful for the computational resources and assistance provided by PROTEUS, the super-computing center of Institute Carlos I in Granada, Spain. We are grateful for access to the University of Nottingham's Augusta HPC service.
\end{acknowledgments}

\bibliography{referencias-BibDesk-OK-v3}{}

\onecolumngrid
\newpage

\renewcommand\thesection{S\arabic{section}}
\renewcommand\theequation{S\arabic{equation}}
\renewcommand\thefigure{S\arabic{figure}}
\setcounter{equation}{0}
\setcounter{figure}{0}

\begin{center}
{\Large SUPPLEMENTAL MATERIAL}
\end{center}
\begin{center}
\vspace{0.8cm}
{\Large Entanglement statistics in Markovian open quantum systems: A matter of mutation and selection}
\end{center}
\begin{center}
Federico Carollo$^{1}$ and Carlos P\'erez-Espigares$^{2,3}$ 
\end{center}
\begin{center}
$^1${\em Institut f\"ur Theoretische Physik, Universit\"at T\"ubingen, Auf der Morgenstelle 14, 72076 T\"ubingen, Germany}\\
$^2${\em Departamento de Electromagnetismo y F\'isica de la Materia, Universidad de Granada, Granada 18071, Spain}
$^3${\em Instituto Carlos I de F\'isica Te\'orica y Computacional, Universidad de Granada, Granada 18071, Spain}
\end{center}

\section*{S1. Quantum Cloning Algorithm} 
\label{app1}
We are interested in the statistics of arbitrary time-averaged observables, ranging e.g.~from the number of photons or particles leaving the system per unit time, to the time-averaged entanglement entropy or coherence up to time $\tau$. By dividing the time interval $[0,\tau]$ into $N$ infinitesimal time-steps $\delta t=\tau/N\to 0$, we can express any time-integrated observable $O_\tau$ as a functional over trajectories $O_\tau={\cal O}[\omega_\tau]$, 
$$
\mathcal{O}[\omega_\tau]=\sum_{k=0}^{N-1}\alpha\left(|\psi_k\rangle,|\psi_{k+1}\rangle\right)\, ;
$$
where $\omega_\tau=\{ |\psi_0\rangle, |\psi_1\rangle, \dots |\psi_{N}\rangle \}$ is the quantum trajectory and $\alpha$  is an observable-dependent function describing the discrete increment of the observable in a time-step. Recall that we are short-hading the pure state at a given time, $\ket{\psi_{k\, \delta t}}$, as $\ket{\psi_k}$.

\noindent In order to study the statistics of such observables we firstly explicit the probability of a quantum trajectory, which by the Markov character of the dynamics of the system can be written as
$$
P[\omega_\tau]=p_{|\psi_{N-1}\ra\to |\psi_{N}\ra}\cdots p_{|\psi_0\ra\to |\psi_1\ra}p_{|\psi_0\ra}\,
$$
where $p_{|\psi_0\ra}$ is the probability of the initial state $|\psi_0\ra$ and $p_{|\psi_k\ra\to |\psi_{k+1}\ra}$ stands for the transition probability from state $|\psi_{k}\ra$ to $|\psi_{k+1}\ra$, which can be either $p_{\mu}=\delta t\bra{\psi_k}J_\mu^{\dagger}J_\mu\ket{\psi_k}$, for $\mu=1,2,...,N_j$ or $p_0=1-\sum_{\mu=1}^{N_j}p_\mu$, depending on whether the system either jumps or evolves under the action of $H_{\rm eff}$, respectively. Here, $N_j$ are the number of jump operators $J_\mu$ encoding the type of interaction between system and environment.
Thus, the probability of having a value $O$ for the observable is,
$$
P_\tau(O)=\sum_{\omega_\tau}P[\omega_\tau]\delta(O-{\cal O}[\omega_\tau])=\sum_{\omega_\tau}p_{|\psi_{N-1}\ra\to |\psi_{N}\ra}\cdots p_{|\psi_0\ra\to |\psi_1\ra}p_{|\psi_0\ra}\delta\left(O-\sum_{k=0}^{N-1} \alpha(|\psi_{k}\ra,|\psi_{k+1}\ra)\right)\, .
$$
The same information is obtained from the moment generating function or dynamical partition function $Z_\tau(s)=\sum_O P(O) e^{-s O}$, which reads 
$$
Z_\tau(s)=\sum_{\omega_\tau}p_{|\psi_{N-1}\ra\to |\psi_{N}\ra}e^{-s \alpha(|\psi_{N-1}\ra,|\psi_N\ra)}\cdots p_{|\psi_0\ra\to |\psi_1\ra}e^{-s \alpha(|\psi_0\ra,|\psi_1\ra)}p_{|\psi_0\ra}\,.
$$
Thus by defining 
\be
{\tilde p}_{|\psi_{k}\ra\to |\psi_{k+1}\ra}\equiv p_{|\psi_{k}\ra\to |\psi_{k+1}\ra}e^{-s \alpha(|\psi_{k}\ra,|\psi_{k+1}\ra)}\,,
\label{ptilde}
\ee
we can rewrite the dynamical partition function as $Z_\tau(s)=\sum_{\omega_\tau}{\tilde p}_{|\psi_{N-1}\ra\to |\psi_{N}\ra}\cdots {\tilde p}_{|\psi_0\ra\to |\psi_1\ra}p_{|\psi_0\ra}$. We are interested in simulating $Z_\tau(s)$  for long times in order to obtain the dynamical free energy, as $\theta(s)=\lim_{\tau\to \infty}\frac{1}{\tau}\ln Z_\tau(s)$.
However, the transition probabilities \eqref{ptilde} are not normalized $(\sum_{|\psi_{k+1}\ra}p_{|\psi_k\ra \to |\psi_{k+1}\ra}\neq 1)$, and cannot be simulated as a standard stochastic process. 
Nevertheless, we can normalize them by introducing the cloning rates, $Y_{|\psi_k \ra}\equiv \sum_{|\psi_{k+1}\ra}{\tilde p}_{|\psi_k\ra \to |\psi_{k+1}\ra}$, so that 
\be
p'_{|\psi_{k}\ra\to |\psi_{k+1}\ra}\equiv \dfrac{{\tilde p}_{|\psi_{k}\ra\to |\psi_{k+1}\ra}}{Y_{|\psi_{k}\ra}}\,,
\label{pprima}
\ee
are proper transition probabilities. We can then express $Z_\tau(s)$ as 
\be
Z_\tau(s)=\sum_{\omega_\tau}p'_{|\psi_{N-1}\ra\to |\psi_{N}\ra}Y_{|\psi_{N-1}\ra}\cdots p'_{|\psi_0\ra\to |\psi_1\ra}Y_{|\psi_0\ra}p_{|\psi_0\ra}\,.
\label{clonpath}
\ee
This way of writing the dynamical partition function $Z_\tau(s)$ allows us to numerically compute $\theta(s)$, since \eqref{clonpath} can be computationally read as follows: each state $|\psi_k\ra$ evolves at each time-step $\delta t$ according to $p'_{|\psi_k\ra\to |\psi_{k+1}\ra}Y_{|\psi_k\ra}$, which consists of a stochastic evolution --or ``{\emph{mutation}}"-- with transition probability $p'_{|\psi_k\ra\to |\psi_{k+1}\ra}$ and a ``{\emph{selection}}" term which replicates or removes the evolved state by replacing it with an integer number of identical copies with average $Y_{|\psi_k\ra}$. The latter operation is the responsible for the exponential growth of $Z_\tau(s)$ in time. To check this we write for given a trajectory $\{|\psi_0\ra,\cdots,|\psi_N\ra\}$, the number of clones at time $\tau$, ${\cal N}(\{|\psi_0\ra,\cdots,|\psi_N\ra\};\tau)$, as the following recurrence relation:
$$
{\cal N}(\{|\psi_0\ra,\dots,|\psi_N\ra\};\tau)=p'_{|\psi_{N-1}\ra \to |\psi_{N}\ra}Y_{|\psi_{N-1}\ra}{\cal N}(\{|\psi_0\ra,\dots,|\psi_{N-1}\ra\};\tau-\delta t)\,,
$$
so iterating we get 
$$
{\cal N}(\{|\psi_0\ra,\dots,|\psi_N\ra\};\tau)=p'_{|\psi_{N-1}\ra \to |\psi_{N}\ra}Y_{|\psi_{N-1}\ra}\cdots p'_{|\psi_0\ra \to |\psi_1\ra}Y_{|\psi_0\ra}{\cal N}(|\psi_0\ra;0)\,,
$$
where ${\cal N}(|\psi_0\ra;0)=N_cp_{|\psi_0\ra}$ is the number of copies in the initial state $|\psi_0\ra$ when starting with a number $N_c$ of initial states. Thus the average number of clones at time $\tau$, $\la {\cal N}(\tau)\ra$ is given by
$$
\la {\cal N}(\tau)\ra=\sum_{\omega_\tau}{\cal N}(\{|\psi_0\ra,\dots,|\psi_N\ra\};\tau)=\sum_{\omega_\tau}p'_{|\psi_{N-1}\ra \to |\psi_{N}\ra}Y_{|\psi_{N-1}\ra}\cdots p'_{|\psi_0\ra \to |\psi_1\ra}Y_{|\psi_0\ra}p_{|\psi_0\ra}=N_cZ_\tau(s)
$$
and consequently $Z_\tau(s)=\la {\cal N}(\tau)\ra/N_c$. However, due to the exponential growth in time of $\la {\cal N}(t) \ra$ --notice that $Z_\tau(s)\sim e^{\tau\theta(s)}$--, it is convenient to write the dynamical partition function as
$$
Z_\tau(s)=\prod_{k=1}^N \frac{\la {\cal N}(k\delta t) \ra}{\la {\cal N}((k-1)\delta t) \ra} \,.
$$
with $\la {\cal N}(0) \ra=N_c$. In order to avoid diverging or vanishing population of clones after a time $\tau$, we resize the population after every time-step to the starting size $N_c$. In this way we can write $Z_\tau(s)=\prod_{k=1}^N \la {\cal \tilde N}(k\delta t) \ra/N_c$, with ${\cal \tilde N}(k\delta t)$ being the number of clones at time $k\delta t$ starting from a population of $N_c$ at time $(k-1)\delta t$. This allows us to compute the dynamical free energy as
$$
\theta(s)\sim\frac{1}{\tau}\sum_{k=1}^N \ln\left( \frac{\la {\cal \tilde N}(k\delta t) \ra}{N_c} \right)\,,~~{\rm for}~~N_c,\tau\gg 1\,.
$$
The computation of the above expression can be numerically performed with the following algorithm at every time-step:
\begin{enumerate}
\item[(a)] \emph{Mutation}: Each of the $N_c$ states $|\psi^i_k\ra$ with $i=1,2,\dots,N_c$, is independently evolved with the stochastic dynamics $p'_{|\psi^i_k\ra \to |\psi^i_{k+1}\ra}$.
\item[(b)] \emph{Selection}: The evolved states are replicated or removed with average $Y_{|\psi^i_k\ra}$. This is carried out by copying the evolved states $y_i=\lfloor Y_{|\psi^i_k\ra} +u\rfloor$ times, with $u$ being a random number uniformly distributed $u\in [0,1]$. If $y_i=0$ then there is no offspring of the evolved state $|\psi^i_{k+1}\ra$.
\item[(c)] \emph{Resizing}: The factor $X_k=N'_c/N_c$ is computed, where $N'_c=\sum_{i=1}^{N_c}y_i$ is the new total number of copies, and the total number of copies is sent back to $N_c$ by adding (if $N'_c<N_c$) or removing (if $N'_c>N_c$) $|N'_c-N_c|$ copies uniformly at random among the $N'_c$ states. 
\item[(d)] Increase the time-step by one $k=k+1$ and go to (a) until time $\tau$ ($k=N$) is reached .
\end{enumerate}
For long times and a large number of clones, $N_c\gg 1$, the factor $X_k$ is a good estimator of $\la {\cal \tilde N}(k\delta t) \ra/N_c$, and the dynamical free energy can be thus computed as 
\be
\theta(s)=\frac{1}{\tau}\sum_{k=1}^N \ln\left(X_k\right)\,.
\label{thetassplit}
\ee
The above algorithm is sketched in Fig.~2(a) of the main text. In addition, from this algorithm it is easy to retrieve the rare trajectories responsible for the rare event just by tracing back the states that have survived until the last time-step, as depicted in Fig.~2(b) of the main text.

\section*{S2. Application to jump-rate observables} 
\label{app2}

The algorithm described above allows us to unveil the statistics of arbitrary time-averaged observables. In the main text, it has been applied to observables which depend on the state, such as the coherence or the entanglement entropy. For this kind of observables, the numerical approach here introduced is the only numerical method available so far to obtain the dynamical free energy. Furthermore, this algorithm can be applied as well to jump-rate observables, i.e.~to observables such as the current or the activity --measuring the number of jumps--, which only take a non-zero value whenever a jump occurs. However, for these observables the dynamical free energy can be obtained in a completely independent manner by means of tilted operator techniques \cite{garrahan10a,carollo18b}. We can thus check the validity of the algorithm by computing the  dynamical free energy for jump-rate observables in different systems.
In the following, we shall unveil the statistics of the activity in a laser-driven two-level atom with decay and the current in a boundary-driven spin chain.

\subsection*{S2.1 Activity statistics in a laser-driven two-level atom with decay}

The activity in this case corresponds to the photon-emission rate of the system. The state of the system can be either excited $\ket{\uparrow}=(1,0)^T$ or de-excited $\ket{\downarrow}=(0,1)^T$. Rabi oscillations are implemented by the Hamiltonian $H=\Omega(\sigma_- + \sigma_+)$, where $\sigma_{-}=\ketbra{\downarrow}{\uparrow}$, $\sigma_{+}=\sigma_-^{\dagger}$, and $\Omega$ the Rabi frequency. The jump operator describing photon-emission is $J_1=\sqrt{\gamma}\sigma_{-}$, with $\gamma$ being the decay rate so that $J_1\ket{\downarrow} =0$ and such that for a state $|\psi\rangle=a \ket{\uparrow} + b \ket{\downarrow}$ with $|a|^2+|b|^2=1$, we have  $J_1|\psi\rangle =\ket{\downarrow}$. To uncover the statistics of the activity, it is possible to exploit tilted operator techniques \cite{garrahan10a,carollo18b}, which show that the dynamical free energy $\theta(s)$ for this observable is given by the largest real eigenvalue of the following tilted Lindbladian 
$$
\mathcal{L}_s[X]=-i[H,X]+e^{-s}J_1XJ_1^\dagger -\frac{1}{2}\left\{X,J_1^\dagger J_1\right\}\, .
$$
On the other hand, within the framework presented here we can obtain the same dynamical free energy applying the quantum cloning algorithm with an increment for the observable $\alpha=1$ if a jump takes place, i.e.~if $|\psi_{k+1}\rangle=J_1|\psi_k\rangle/\|J_1|\psi_k\rangle\|$,
and zero otherwise. 

\begin{figure}[h!]
\includegraphics[scale=0.6]{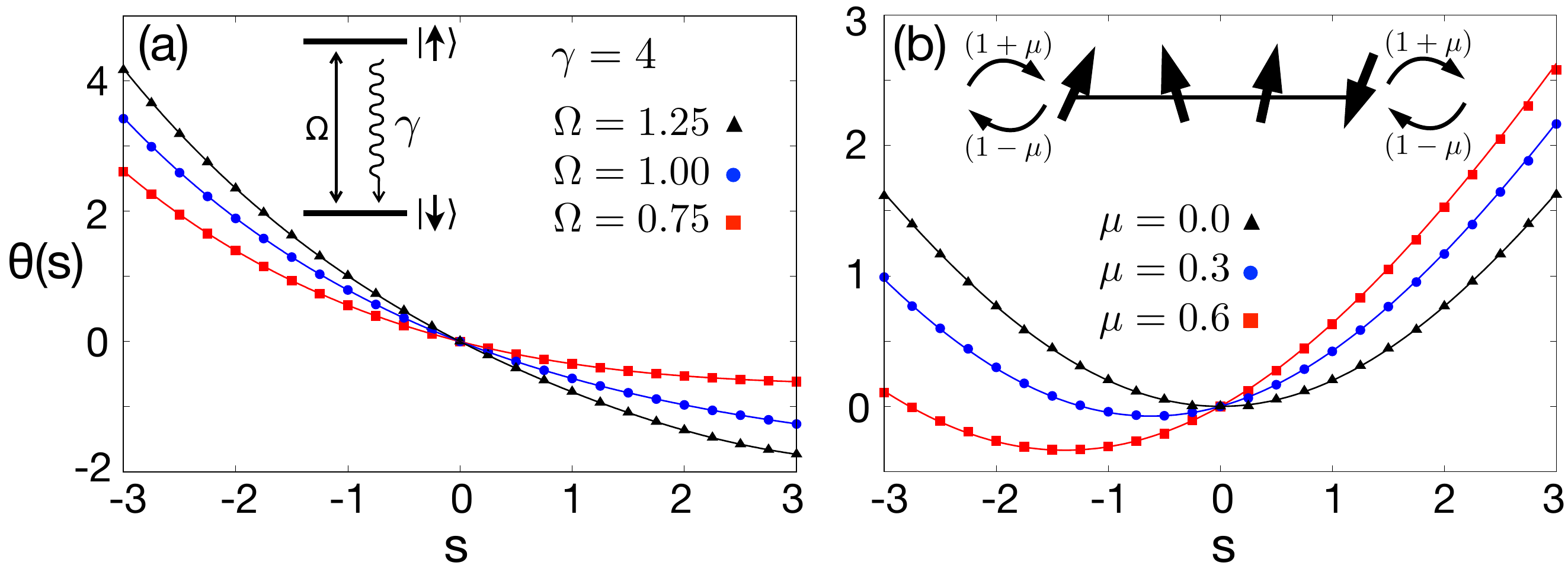}
\caption{(a) Dynamical free energy $\theta(s)$ for the time-averaged activity of a two-level system. Solid lines correspond to exact diagonalisation results while points are the values obtained with the quantum cloning algorithm. (b) Same results for time-averaged current in the boundary-driven XX spin chain for $L=4$ spins.
}
\label{figS1}
\end{figure}

Results are displayed in Fig.~\ref{figS1}(a), where we observe a perfect agreement between the exact diagonalisation prediction (solid lines), and the data obtained with the algorithm (points). In this case the data have been generated for $\gamma=4$ and different Rabi frequencies, with $N_c=5000$ clones and a discretisation step of $\delta t=0.0025$ for trajectories of length $\tau=50$ after a relaxation time of $\tau_{\rm relax}=25$. We then have computed $\theta(s)$ taking the average over $100$ different numerical experiments (or trajectories). Further, the initial state of each clone has been randomly taken.

\subsection*{S2.2 Current statistics in a boundary-driven chain}

Another paradigmatic jump-rate observable besides the activity is the current. Here we focus on the statistics of the current in the boundary-driven XX chain, which is a paradigmatic nonequilibrium system which allow for the study of transport properties in quantum Hamiltonian systems connected at their extreme sites to  particle reservoirs. The effects of these reservoirs consist in injecting or ejecting particles from the bulk of the system. Here we consider sites made of two-level subsystems. Considering $L$ sites, we can write with the notation $x^{(k)}$ the operator $x$ of site $k$, with $x$ being an operator of the spin-$1/2$ algebra. This type of systems are usually modelled, in the Markovian approximation, by Lindblad generators of the following form
\begin{equation}
\begin{split}
\mathcal{L}[X]&=-i[H,X]\\
&+\sum_{\beta=+,-}\gamma_\beta\left(\sigma_\beta^{(1)}X\sigma_\beta^{(1)\dagger}-\frac{1}{2}\left\{X,\sigma_\beta^{(1)\dagger}\sigma_\beta^{(1)}\right\}\right)\\
&+\sum_{\beta=+,-}\gamma_{-\beta}\left(\sigma_\beta^{(L)}X\sigma_\beta^{(L)\dagger}-\frac{1}{2}\left\{X,\sigma_\beta^{(L)\dagger}\sigma_\beta^{(L)}\right\}\right)\, ,
\end{split}
\label{M:chains}
\end{equation}
where $\sigma_{-}=\ketbra{\downarrow}{\uparrow}$ describes ejection of a particle and $\sigma_+=\sigma_-^\dagger$ injection. Rates are given by $\gamma_{+}=1+\mu$ and $\gamma_{-}=1-\mu$, where $\mu$ is the driving parameter establishing an asymmetry between the left and the right boundary. We consider a tight-binding Hamiltonian 
$$
H=\sum_{k=1}^{L-1}\left(\sigma_-^{(k)}\sigma_+^{(k+1)}+\sigma_+^{(k)}\sigma_-^{(k+1)}\right)\, .
$$
In a boundary-driven chain the current corresponds to the net number of particle leaving the chain from, say, the right boundary, per unit time. Again, the statistics in this case can be recovered by obtaining the dynamical free energy $\theta(s)$ as the largest real eigenvalue of the tilted operator \cite{garrahan10a}
\begin{equation*}
\begin{split}
\mathcal{L}_s[X]&=\mathcal{L}[X]+\gamma_+(e^s-1)\sigma_-^{(L)}X\sigma_+^{(L)}\\&+\gamma_-(e^{-s}-1)\sigma_+^{(L)}X\sigma_-^{(L)}
\end{split}\, .
\end{equation*}
With the algorithm here introduced we can derive the same statistics by exploiting the framework described in the main text with observable increments $\alpha$ such that $\alpha=1$ if a particle is removed from the last site --right boundary--, i.e.
$$
|\psi_{k+1}\rangle=\frac{\sigma_-^{(L)}|\psi_k\rangle}{\|\sigma_-^{(L)}|\psi_k\rangle\|}\, ,
$$
or $\alpha=-1$ if a particle is injected in the last site, i.e.
$$
|\psi_{k+1}\rangle=\frac{\sigma_+^{(L)}|\psi_k\rangle}{\|\sigma_+^{(L)}|\psi_k\rangle\|}\, ,
$$
and zero otherwise. 

In Fig.~\ref{figS1}(b) we observe the good agreement between the exact diagonalisation results (solid lines) and the numerical data obtained with the algorithm (points). These data have been generated for a chain of $L=4$ sites with different values of the boundary driving $\mu$, with $N_c=5000$ clones and a discretisation step of $\delta t=0.005$ for trajectories of length $\tau=50$ after a relaxation time of $\tau_{\rm relax}=25$. The dynamical free energy  $\theta(s)$ has been computed taking the average over $100$ different numerical experiments (or trajectories). The initial state of each clone has been randomly taken.

\newpage

\section*{S3. Coherence statistics in the laser-driven two-level atom with decay}
To model the  laser-driven two-level atom emitting photons in an environment we exploit a spin-$1/2$ algebra. The state of the system can be either excited $\ket{\uparrow}=(1,0)^T$ or de-excited $\ket{\downarrow}=(0,1)^T$. Rabi oscillations are implemented by the Hamiltonian $H=\Omega\left(\sigma_-+\sigma_+\right)$, where $\sigma_{-}=\ketbra{\downarrow}{\uparrow}$,  $\sigma_+=\sigma_-^\dagger$, and $\Omega$ the Rabi frequency. The jump operator describing photon-emission is $J_1=\sqrt{\gamma}\sigma_{-}$, with $\gamma$ being the decay rate. The jump operator $J_1$ thus acts as $J_1|\psi\rangle \propto\ket{\downarrow}$ and $J_1\ket{\downarrow} =0$.

In order to derive the dynamical free energy $\theta(s)$ for the coherence, measuring the amount of superposition between the two classical states $\ket{\downarrow}$ and $\ket{\uparrow}$, we firstly write the quantum pure state, at each (discrete) time of the trajectory, as 

\begin{equation}
|\psi_k\rangle =\begin{pmatrix}
a_k\\b_k
\end{pmatrix}\, ,
\label{M:psi}
\end{equation}
with $|a_k|^2+|b_k^2|=1$. Recall that $|\psi_k \rangle$ is short-hand for $|\psi_{t=k\, \delta t}\rangle$. In a density matrix formulation the same state is written as 

$$
|\psi_k\rangle \langle \psi_k|=\begin{pmatrix}
|a_k^2|&a_kb_k^*\\
a_k^*b_k&|b_k|^2
\end{pmatrix}\, ;
$$ 
while diagonal entries represent the population in the classical configuration states $\ket{\uparrow}$ and $\ket{\downarrow}$, the off-diagonal term $c_k=|a_kb_k^*|=|a_k^*b_k|=|\bra{\uparrow}\psi_k\rangle \langle \psi_k\ket{\downarrow}|$ measures  the quantum coherence, i.e.~how much the quantum state is in a superposition, {\it c.f.} Eq.~\eqref{M:psi}. For the statistics of coherence we thus exploit the cloning algorithm with $f=|ab^*|$. The data of Fig.~3(a) have been generated for $\Omega=1$, $\gamma=4$, with $N_c=1000$ clones and a discretisation step of $\delta t=0.01$ for trajectories of length $\tau=500$. We then have computed $\theta(s)$ taking the average over $100$ different numerical experiments (or trajectories). Further, the initial state of each clone has been set to $\ket{\psi_0}=\ket{\downarrow}$. The inset to Fig.~3(a) displaying $\tau^{-1}\ln P_{\tau}(C)$ has been computed by Legendre transforming $\theta(s)$.

\section*{S4. Entanglement statistics in boundary-driven chains}
%
%

In this case, the Lindblad generators have the same form of \eqref{M:chains}. Rates are given by $\gamma_{+}=1+\mu$ and $\gamma_{-}=1-\mu$, where $\mu$ is the driving parameter establishing an asymmetry between the left and the right boundary. We consider a tight-binding Hamiltonian 

$$
H=\sum_{k=1}^{L-1}\left(\sigma_-^{(k)}\sigma_+^{(k+1)}+\sigma_+^{(k)}\sigma_-^{(k+1)}\right)+\Delta\sum_{k=1}^{L-1}\sigma_z^{(k)}\sigma_z^{(k+1)}\, ,
$$
where $\sigma_z=\ketbra{\uparrow}{\uparrow}-\ketbra{\downarrow}{\downarrow}$ is the single-site magnetisation operator. For non-interacting systems one sets $\Delta=0$ and the problem can be mapped into a free-fermionic model. The presence of strong interactions is instead accounted for by a $\Delta\gg1$.

As a quantification of bipartite entanglement in the stochastic pure state of the chain, we consider the Von Neumann entropy of the reduced density matrix of one of the two subsystems. Dividing the system into two halves of length $L/2$, which we call $A$ and $B$, we evaluate the reduced density matrix of the subsystem $A$ as 

$$
\rho_A=\Tr_B\left(|\psi\rangle \langle \psi|\right)\, ,
$$ 
where $|\psi\rangle$ is the pure state of the full system and $\Tr_B$ denotes trace over the degrees of freedom in $B$. The Von Neumann entanglement entropy of the bipartition is thus defined as 

$$
S_{VN}=-\Tr\left(\rho_A\log \rho_A\right)\, .
$$
Therefore, in order to unveil the statistics of the entropy of entanglement we consider, for the framework described in the main text, an increment 

$$
\alpha(|\psi_k\rangle, |\psi_{k+1}\rangle)=\frac{\delta t}{2}\left[f(|\psi_k\rangle)+f(|\psi_{k+1}\rangle)\right]\, ,
$$
with $f=S_{VN}$. 

The data of Fig.~3(b) have been generated for a spin chain with $L=2$ sites, $\mu=0.5$, $\Delta=0$, using $N_c=1000$ clones for trajectories of length $\tau=50$ with a $\delta t=0.01$ after a relaxation time of $\tau_{\rm relax}=25$. The dynamical free energy $\theta(s)$ has been computed by taking the average over $100$ different numerical experiments. In this case, we have observed that assigning to each clone a  random initial condition will most likely give rise to a clone with a highly entangled initial state ($S_{VN}\approx \log(2)$). Such clone will then be favoured by the dynamics of the algorithm and will avoid jumping for sufficient large negative $s$ ($s\lesssim -5$). This happens because once the favourable dynamical regime, generated by the effective Hamiltonian dynamics of the initial state, is left, it cannot be obtained again during the stochastic process. This reflects, as highlighted in the main text, how quantum jump processes --unlike commonly studied classical systems-- do not visit the whole space of admissible pure states. For that reason we initially set every clone in the classical configuration $\ket{\psi_0}=\ket{\uparrow \downarrow}$. Finally, the inset to Fig.~3(b) displaying $\tau^{-1}\ln P_{\tau}(E)$ has been computed by Legendre transforming $\theta(s)$.

The data of Fig.~4 have been generated for a spin chain with $L=4$ sites, maximum driving $\mu=1$, and strong interaction $\Delta=4$. In this case we have used $N_c=1000$ clones for trajectories of length $\tau=50$ with a $\delta t=0.05$ after a relaxation time of $\tau_{\rm relax}=25$. The dynamical free energy $\theta(s)$ has been computed by taking the average over $100$ different numerical experiments for  random initial conditions. The inset to Fig.~4(a) displays the hysteresis loop obtained by starting with the different initial conditions. These data have been obtained by time-averaging (for intermediate times, i.e.~, between $\tau_{\rm ini}=10$ and $\tau_{\rm end}=40$ to avoid time-boundary effects \cite{garrahan09a}) the entropy of entanglement of each trajectory generated for every clone in each numerical experiment. It is worth noting that in order to obtain the correct values of $\theta(s)$ for $0< s \leq-0.2$ we have to take $\delta t \geq 0.05$, otherwise the product $s\delta t$ in the argument of the exponential of the cloning rate $Y_{\ket{\psi_k}}$ is so small that $Y_{\ket{\psi_k}}=1$, so that no cloning process takes place. In particular this is solved by taking $\delta t=0.1$ for $s=-0.2,-0.15$, $\delta t=0.2$ for $s=-0.10$ and $\delta t=0.3$ for $s=-0.05$.

\section*{S5. Preliminary results about quantum cloning algorithm with matrix product states.}
An interesting application of the quantum cloning algorithm involves the study of genuine nonequilibrium phase transitions in trajectory space for \emph{many-body} quantum systems. In such case, however, it becomes --for increasing system sizes-- rapidly impossible to represent exactly the quantum state on a computer due to exponential growth of the number of parameters contained in the wave function.

Therefore, in order to exploit the quantum cloning for many-body systems it is necessary to combine such algorithm with approximate representations of quantum states. In particular, for one-dimensional systems it is possible to exploit matrix product states (MPSs), together with all the algorithms devised for their time-evolution. In practice, what one can do is to represent quantum trajectories with MPSs as done for instance in Ref.~\cite{Gillman_2019}. Each clone is thus represented via an MPS approximation of its state and the biased time-evolution can be performed, for instance, via a simple time-evolving block-decimation algorithm \cite{PAECKEL2019167998}. On top of such dynamics for each clone one then applies the quantum cloning algorithm. When a clone is replicated one replicates its MPS while when a clone is pruned its MPS must be erased. 

We apply here this strategy to an open many-body system with dynamics described by the Hamiltonian 
\begin{equation}
H=\Omega\sum_{k=1}^L \sigma_x^{(k)}+V\sum_{k=1}^{L-1}n^{(k)}n^{(k+1)}
\label{H-MPS}
\end{equation}
and by the dissipator
\begin{equation}
\mathcal{D}[\rho]=\gamma \sum_{k=1}^L\left(\sigma_-^{(k)}\rho\sigma_+^{(k)}-\frac{1}{2}\left\{n^{(k)},\rho\right\}\right)\, ,
\label{D-MPS}
\end{equation}
describing the possibility of excitation emission from each site. We notice that this model has an extensive number of jump operators as opposed to the boudary driven models previously discussed and presented in the main text. The dynamical free energy $\theta(s)$ for the total emissions in this model can be obtained, via numerical exact diagonalization, as the largest real eigenvalue of the tilted operator
$$
\mathcal{L}_s[\rho]=-i[H,\rho]+\gamma \sum_{k=1}^L\left(e^{-s}\sigma_-^{(k)}\rho\sigma_+^{(k)}-\frac{1}{2}\left\{n^{(k)},\rho\right\}\right)\, .
$$
Notice that positive (negative) values of $s$ hinder (enhance) the activity, i.e. the number of emitted excitations, with respect to the unbiased dynamics, corresponding to $s=0$.

\begin{figure}[t]
\includegraphics[scale=0.48]{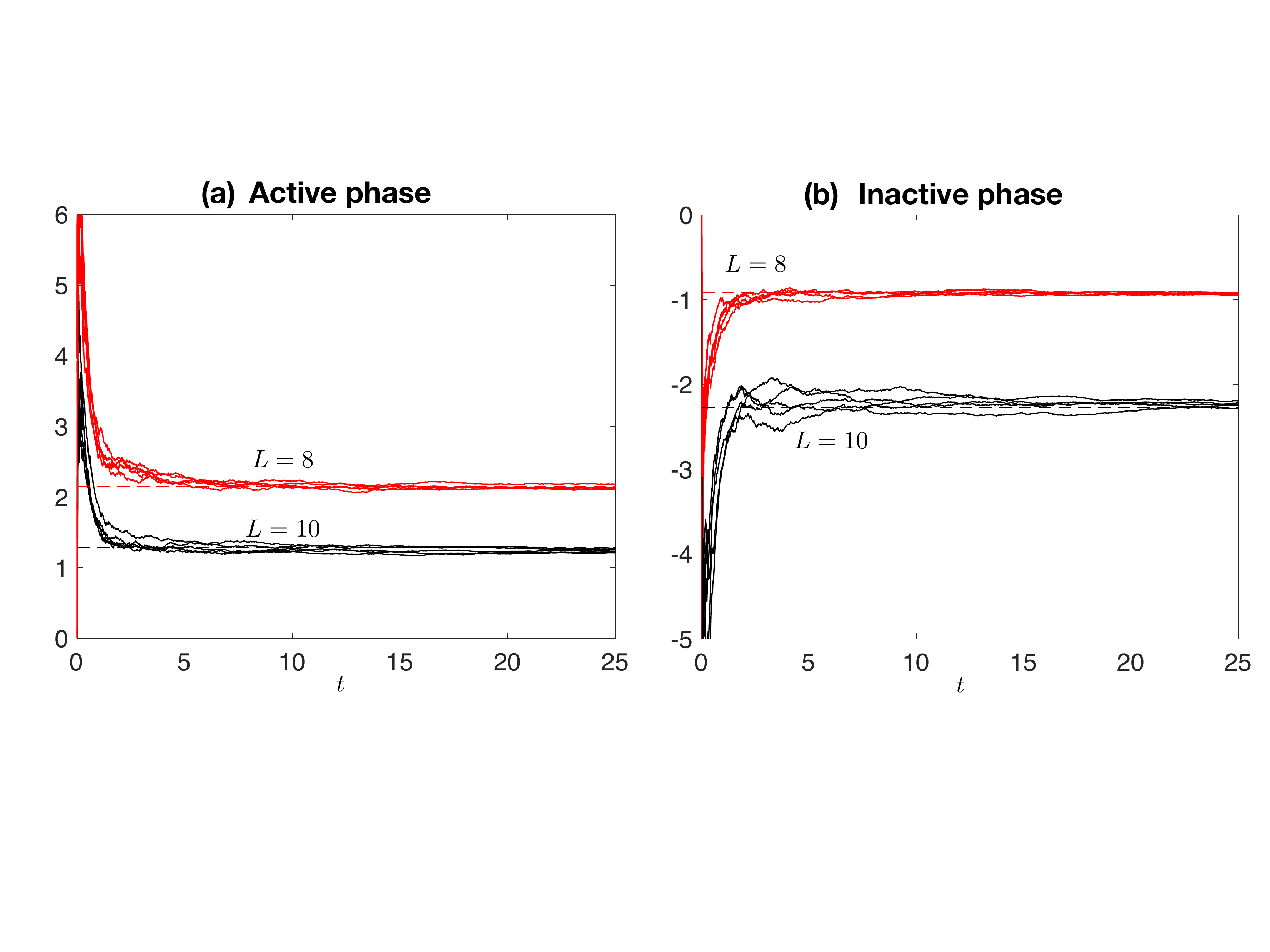}
\caption{ Dynamical free energy $\theta(s)$ for the model described by Eq.~\eqref{H-MPS},\eqref{D-MPS}. Dashed lines represent the exact solution from exact diagonalization of the corresponding tilted operator. Solid lines are the results for several numerical experiments applying the quantum cloning algorithm. We set $\gamma=3$, $\Omega=3/4$ and we take bond dimension $\chi=16$ to approximate the state of the  system in quantum trajectories with MPS. We explore results from a population of $50$ clones and different $dt\in[0.005; 0.02]$ (a) Active phase: negative $s$. We show results for $L=8$ (red lines) with $s=0.25$ and $V=1$, as well as for $L=10$ (black lines) with $s=0.5$ and $V=0.5$. (b) Inactive phase: positive $s$. We report results for $L=8$ (red lines) with $s=-0.5$ and for $L=10$ (black lines) with $s=-0.25$, in both cases setting $V=0.5$.}
\label{figS2}
\end{figure}

In Fig.~\ref{figS2}, we provide results for the dynamical free energy as obtained through the application of the quantum cloning algorithm combined with MPS. The dashed lines correspond to the exact values of the dynamical free energy for the different parameters, while solid lines represent the dynamical free energy (as a function of simulation time) resulting from the quantum cloning algorithm exploiting MPSs. Each line is representative of a numerical experiment of the evolution of the population of clones. As apparent from Fig.~\ref{figS2}, the agreement is remarkable already with a rather small number of clones. These results have been obtained on a personal computer and simulation time required was of the order of few hours per experiment. This shows how applying the quantum cloning in combination with MPS can lead to a very powerful strategy to investigate dynamical behaviour of many-body quantum systems. 

\end{document}